# Radiative-field quantum-coupling between closely-spaced surfaces

K. P. Sinha[a] and A. Meulenberg[b]

[a] Department of Physics, IISc, Bangalore 560012, India (email: kpsinha@gmail.com)
[b] Science for Humanity Trust, Inc., Tucker, GA, USA (email: mules333@gmail.com)

A quantum-mechanical formulation of energy transfer between closely-spaced surfaces is given. Coupling between the two surfaces arises from the atomic dipole-dipole interaction involving transverse-photon exchange. The exchange of photons at resonance greatly enhances the radiation transfer. The spacing (distance) dependence is derived for the quantum well - quantum well situation. The interaction between two planar quantum wells, separated by a gap is found to be proportional to the 4$^{th}$ power of the wavelength-to-gapwidth ratio and to the radiation tunneling factor for the evanescent waves. Expressions for the net power transfer, in the near-field regime, from hot to cold surface for this case is given and evaluated for representative materials. Computational modeling of selected, but realizable, emitter and detector structures and materials shows the benefits of both near-field and resonance coupling (e.g., with 0.1 μm gaps).

INTRODUCTION

Thermophotovoltaics (TPV), which converts light (more accurately thermally-generated, long-wavelength-infrared radiation) from a heated surface into electricity, is beginning to grow as a result of new materials capabilities. Specifically, new semiconductor materials allow the photoconverters to convert long-wavelength light into electrical power more efficiently by providing a narrower electrical band gap, better generation of photo-excited minority carriers and their collection at the p-n junction, and reduced recombination dark-current, which controls the open-circuit voltage of the devices. However, there are some approaches, based on improved technical capabilities, which can compound all of these improvements and, potentially, allow TPV to become even more useful.

A requirement of a TPV system is a thermal difference between the emitter and PV device. This means that thermal isolation, in the form of distance (or of a vacuum, if the distance is small), must separate the devices. However, it is known that elimination of the gap can greatly enhance the transfer of radiation between them. The ability to increase the transfer of optical photons across a very small gap (a fraction of a wavelength) without allowing heat flow, via phonons, has been demonstrated.[1,2] A theoretical basis for the enhanced optical transfer has been laid in a quantum mechanical framework.[3] Since this early work, there have been more developments in this area.[4,5] Some review papers [6,7] and even a book[8] have been published with material on the topic.

A number of years ago, following high-temperature (>900C) IV measurements on small (2x2mm) individual InAs photocell samples and 2x2 arrays,[2] it was demonstrated (but not published) that, even for practical-size devices (2x2 cm) at high temperatures (>700C), the optical throughput can be much-enhanced with the use of a very-narrow vacuum gaps (≥ 0.2 μm). This enhancement, through the use of a "submicron" gap, is the basis for Microgap ThermoPhotoVoltaics (MTPV).[9,10] Despite confirmation of the theoretical concept, and proof of its technical feasibility, little work has progressed in the implementation of this technology other than that conducted by a startup company (MTPV Corp.),[11] which grew out of this foundational work. Since the review papers, have not covered such experimental work and have not indicated some of the optimal paths for successfully implementing the MTPV process, we believe that an additional paper on the subject might be warranted. We include some of the remarkable predictions for realizable products.

One enhancement mechanism, available to MTPV and demonstrated in Ref. 1, is purely a physical-optics effect. It involves the development of an "effective" refractive index ($n_g$) within the microgap. As the gap approaches zero width, its effective refractive index $n_g$ approaches that of the emitter and PV device (assume $n_e = n_{pv}$ for optimum results in MTPV). As the difference between refractive indices of the devices and the gap diminishes, the critical angle of total internal reflection (TIR) increases and the percentage of thermally-generated light that can escape from the emitter increases. As the gap width approaches zero, the enhancement from this mechanism approaches a maximum of $n_e^2$. This effect (called the $n^2$ effect) is a macroscopic effect and is independent of the distance between <u>individual</u> atomic radiators (excited atoms or dipoles) in the emitter and absorbers (ground-state atoms) in the PV device. The $n^2$ effect depends only on the gap between the emitter and detector. However, the primary subject of this paper is enhancement of the optical coupling between atoms that <u>is</u> dependent on the distance between <u>individual</u> radiators and absorbers. This process will be called "resonance-enhancement." ("Proximity-enhancement" includes both the macroscopic $n_e^2$ effect and the atomic-level resonant-enhancement effect.)

Resonance enhancement, a "beyond $n_e^2$ effect," involves the interaction of dipole oscillators that are close enough together to be "coupled" by another oscillator, the photon.[12] This "correlated" interaction provides enhanced optical coupling and may be compared with the non-correlated interaction of independent creation and absorption of photons. Closer dipoles (in space and in transition energy or frequency) give stronger coupling. Thus, the phenomenon is both a proximity effect and a resonance effect. While other mechanisms have been developed to explain the enhanced optical-coupling effects across a microgap,[6,7] the mathematical descriptions are sometimes the same. Since surface proximity of the source and detector is the strongest effect, we feel that planar-geometry techniques (e.g., selective-emission and surface-coupling films) are likely to be the most productive. The ability to combine multiple effects in the surface layers is critical to the success of a structure. On the other hand, some of these same techniques can be used in improving thermal emitters to be used in far-field applications. In this latter application, surface coupling is not important and some of the material and structural requirements can be relaxed. Different techniques might be optimal.

In the following, the derived expression will explicitly contain the three effects, namely, the gapwidth (source/absorber-separation) dependence, the $n^2$ enhancement, and the resonance enhancement. The quantum-coupling approach presented here, in the context of small-gap TVP converters, had not been considered previously for macroscopic devices. There do exist, however, relevant papers on the radiation field and fluctuations in close proximity to a hot emitter, for example, those based on a classical radiation-intensity approach[13] and others based on a classical formulation of fluctuation electrodynamics.[14] Most of the published studies[15,16,17,18,19], follow the latter approach in which the randomly fluctuating charge and current densities provide a source of radiation.

THEORETICAL MODEL

In the current model, we consider two-level systems between separated dielectrics at different temperatures, as shown in Figure 1. The two slabs are separated by a dielectric gap[20] of width $\ell$, which can be varied. While our discussion has focused above on the case of coupling across a vacuum gap, it is possible that the first experiments to confirm the model for quantum-well / quantum-well coupling will be done with a thin solid gap made of an insulator, and the thermal excitation of the emitter will be replaced by an optical excitation scheme. Various parameters of the regions, such as temperatures, $T_1$ ($\gg T_2$), dielectric functions, $\varepsilon_i(\omega)$ - where $\omega$ is the frequency - will be denoted by the corresponding suffixes.

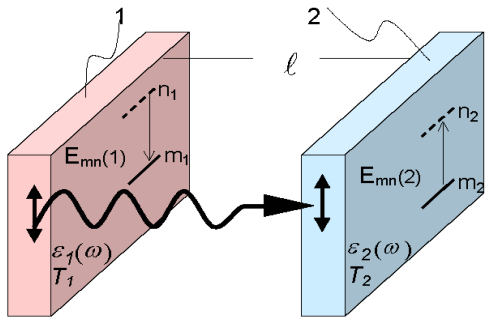

*Figure 1. Coupling between two-level quantum systems separated by a gap.*

We recognize that there may be many two-level systems on both the emitter side and on the absorber side of the gap. We will assume that the coupling between the two-level systems is radiative, including contributions from both electrostatic (Coulomb) and electrodynamic fields. This implies an underlying Hamiltonian of the form

$$\hat{H} = \sum_i \hat{H}_{atom}^{(1)}[i] + \sum_j \hat{H}_{atom}^{(2)}[j] + \hat{H}_{rad} + \hat{H}_{coul} + V_{dipole} \tag{1}$$

In this equation, the first summation is over emitter-side two-level atomic models indexed by *i*. The second summation is over the absorber-side two-level atomic models, indexed by *j*. The transverse radiation field is denoted through $\hat{H}_{rad}$, and the Coulomb interaction is included through $\hat{H}_{coul}$. The model involving the longitudinal electrodynamic mode of $H_{Coul}$ (included in the evanescent waves)[21] had been discussed by the authors and is mentioned elsewhere.[22] In what follows, we consider coupling through $V_{dipole}$, which involves the transverse electric component of the Coulomb source field.

The atomic transitions are presumed to be dominated by the electric-dipole transitions. Thus, we may extract the dipole-dipole interaction from a multipole expansion of the interaction:

$$V_{dipole} = -\sum \boldsymbol{\mu}_j \cdot \mathbf{E}_T(\mathbf{r}_j) \tag{2}$$

where **μ** is the dipole-moment operator and $\mathbf{E}_T$ is the dipole (transverse-) electric-field operator.

Explicitly in the case of two dipoles in free space, the interaction Hamiltonian, $H_{int} = \boldsymbol{\mu}(1) \cdot \mathbf{E}_{T2}(R_1) - \boldsymbol{\mu}(2) \cdot \mathbf{E}_{T1}(R_2)$, where $\mathbf{E}_{T2}(R_1)$ and $\mathbf{E}_{T1}(R_2)$ are the Coulomb source fields (field at $R_1$ from dipole 2 and field at $R_2$ from dipole 1, respectively). The coupling between a pair of atoms (dipoles), on using the above interaction, will appear in fourth-order perturbation theory. However, it is convenient to transform the Hamiltonian by using a unitary transformation,[23,24,25,26] $H_t = e^{iS} H e^{-iS}$, where S is an operator chosen to eliminate the linear term **μ** · $\mathbf{E}_T$. Then, the effective interaction between two dipoles is obtained in the second-order perturbation theory.

Averaging over all polarizations and angles leads to the effective interaction between two <u>randomly-oriented</u> dipoles at a distance $R_{ij}$ apart in free space. For the near-field (NF) limiting form, this is:

$$\left[U^{NF}(R_{ij})\right]^{(2)} \approx -\frac{2}{3} \frac{|\mu_{mn}(2)|^2 |\mu_{mn}(1)|^2}{R_{ij}^6 \Delta E_{12}} \tag{3}$$

where $\Delta E_{12}$, is positive and $\mu_{mn}$ are the matrix elements of the transition-dipole moments. In the near-field region, the dipoles are separated by much less than a wavelength at the energy of interest, $R_{ij} \ll \lambda$. In the far-field region, $R_{ij} \gg \lambda$.

For real-photon (or resonant) exchange, one side must be in the excited state and the other must be in the ground state. Thus, $\Delta E_{12}(r) = \hbar(\omega_2 - \omega_1)$, where $\omega_2$ and $\omega_1$ are frequencies associated with the respective transitions (Figure 2).

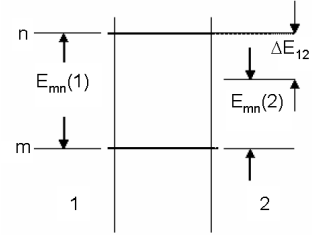

*Figure 2. Schematic representation of the two-medium, two-level system used in the model.*

For off-resonant (o-r) photon exchange, $\Delta E_{12}$(o-r) is not equal to zero and the off-resonant result is found from the Cauchy Principal Value of the integral. For resonant-photon exchange, $\Delta E_{12} = \left\| E_{mn}(1) \right| - \left| E_{mn}(2) \right\|$ in the denominator of Equ. (3) is vanishingly small. Hence, that equation must be solved with special care via the Method of Residues or use of the delta function, $\delta(\omega - \omega_0)$. Generally, the off-resonant terms are found to be small relative to the resonant terms.

For the London interaction,[24] the term in the denominator of Eq. (3) is $E_{mn}(1) + E_{mn}(2)$ rather than $\Delta E_{12}$. Since $E_{mn}(1) + E_{mn}(2) \gg \Delta E_{12} = \left\| E_{mn}(1) \right| - \left| E_{mn}(2) \right\|$, any energy exchange via this pathway is negligible relative to real photon exchange.

In the system considered here, the absorbers are anchored in a medium, and hence, the complex dielectric constants of the material medium ($\varepsilon_i(\omega) = \varepsilon_i'(\omega) + i\varepsilon_i''(\omega)$) are involved and often appear in the corresponding equations (e.g., Van Kampen et al.[27] and Dzyaloshinskii et al.[28]). Although the dipoles are embedded in a dielectric, the transition matrix elements of the dipole moment are estimated from the oscillator strengths[29] (which are proportional to the absorption coefficients, as described by the imaginary portion of the dielectric function, or proportional to the square of the dipole matrix elements). In our case, we are using the dipole moments derived from the oscillator strengths, and therefore, Eq. (3) (with dielectric medium effects using the method of images) becomes:

$$[U^{NF}(R_{ij})]^{(2)} \approx -\frac{2}{3} \frac{16\,\varepsilon_o(\omega)^2 |\mu_{mn}(2)|^2 |\mu_{mn}(1)|^2}{R_{ij}^6 |\varepsilon_1(\omega)+\varepsilon_o(\omega)|^2 |\varepsilon_2(\omega)+\varepsilon_o(\omega)|^2 \Delta E_{12}} \tag{4}$$

The above formulation is for a pair of atoms (each a dipole), one on side 1 and another on side 2. The effect of <u>all</u> atoms on both sides can be obtained by integrations over both half spaces (see Ref. 25). If the single atom on side 1 is replaced by half space $Z \leq 0$ with $N_1$ atoms per unit volume (p.u.v.) and the other at half space $Z \geq \ell$ (where $\ell$ is the distance between the two slabs) with $N_2$ atoms (p.u.v.), the NF (= near field), interaction energy per unit area for the current situation is:

$$[U^{NF}(\ell)]^{(2)} = -\frac{4}{9}\frac{(2\pi)\,N_1 N_2\,\varepsilon_o(\omega)^2}{|\varepsilon_1(\omega)+\varepsilon_o(\omega)|^2 |\varepsilon_2(\omega)+\varepsilon_o(\omega)|^2} \frac{|\mu_{mn}(2)|^2 |\mu_{mn}(1)|^2}{\Delta E_{12}} G(\ell), \tag{5}$$

where $G(\ell) = 1/\ell^2$. We have derived the geometric term, $G(\ell)$ for two quantum wells of width $W_{qw}$, separated by a vacuum gap $\ell$.

$$G = \frac{1}{\ell^2}\left[1 - \frac{2}{(1+(W_{qw}/\ell))^2} + \frac{1}{(1+(2W_{qw}/\ell))^2}\right] \tag{6}$$

A real, transverse photon is created on the hot side and absorbed in the cold side. As we are considering very short distances here (NF), the evanescent[21] mode will dominate the radiation transfer in this region. Within the dipole-dipole interaction via the radiation field, the transfer of energy from slab 1 to slab 2 and the reverse process is to be considered. Before writing the full expression, we must note the following situations.

The probability of emission of a quantum of radiation depends on the factor $[n(\omega, T_1)+1]$ and the absorption on the other side depends on $n(\omega, T_2)$, where $n(\omega, T) = [1/\{\exp(\hbar\omega/K_B T) - 1\}]$, the Bose distribution function, $F(B)$. The expression will involve:

$$[n(\omega, T_1)+1][n(\omega, T_2)] - [n(\omega, T_2)+1][n(\omega, T_1)] = -[n(\omega, T_1) - n(\omega, T_2)] \tag{7}$$

For case 1, with a quantum well on each side of the gap (areal structures), the net power transfer (aa => area-to-area) is the photon energy times the transition probability of photon transfer, $\Gamma = \Gamma_{12} - \Gamma_{21}$

$$P_{NF}(aa) = \int \hbar\omega\,\Gamma\,d\omega = \frac{2\pi}{\hbar}\int |U_{NF}(\omega,\bar{\ell})|^{(2)} \rho^\alpha(\omega)[n(\omega,T_1) - n(\omega,T_2)](\hbar\omega)\,d\omega \tag{8}$$

where $\rho^\alpha(\omega)$ is the areal density of states and $(\bar{\ell})$ is the dimensionless version of the gap width. (Explicitly, $(\bar{\ell}) = \ell\, n_g\,\omega/2\pi c$). Equation (8) may be expanded into:

$$P_{NF}(aa) = \frac{2\pi}{\hbar}\int (8\pi/9)\frac{|\mu_{mn}^{(1)}|^2 |\mu_{mn}^{(2)}|^2 N_1^a N_2^a}{|\hbar(\omega-\omega_o)|}(\hbar\omega)4\pi\left(\frac{\omega\,n_e^2}{c^2}\right)\left(\frac{\omega^2\,n_g^2}{(2\pi c)^2}\right)$$
$$* [n(\omega,T_1) - n(\omega,T_2)]\,F(M)\,G(\bar{\ell})T_{12}(\bar{\ell})\,d\omega \tag{8a}$$

$$= \left(\frac{2\pi}{\hbar}\right)\int\left(\frac{8\pi}{9}\right)|\mu_{mn}^{(1)}|^2 |\mu_{mn}^{(2)}|^2\,N_1^a N_2^a\,\delta(\omega - \omega_o)\left(\frac{\omega^4 n_i^4}{\pi c^4}\right)$$
$$* [n(\omega,T_1) - n(\omega,T_2)]\,F(M)\,G(\bar{\ell})T_{12}(\bar{\ell})\,d\omega \tag{8b}$$

where $N_1^a$, $N_2^a$ are the number of dipoles per unit area on either side of the gap. Since, for very small gap size, the system resembles an effective uniform dielectric rather than an interrupted dielectric, we use $n_i$ for the (constant) refractive indices of the two materials and gap in question.

Integration over $\omega$ with use of the delta function leads to:

$$P_{NF}(aa) = \left(\frac{2\pi}{\hbar}\right)\left(\frac{8\pi}{9}\right)|\mu_{mn}(1)|^2 \, |\mu_{mn}(2)|^2 \, N_1^a \, N_2^a \left(\frac{\omega_o^4 \, n_i^4}{\pi \, c^4}\right)$$

$$* \, [n(\omega_o, T_1) - n(\omega_o, T_2)] \, T_{12}(\bar{\ell}) \, G(\bar{\ell}) \, F(M))$$

$$= (2\pi/\hbar) \, F(D) \, F(B) \, F(M) \, G(\bar{\ell}) \, F(\omega_o) \, T_{12}(\bar{\ell}) \qquad (9)$$

The factor, $F(\omega_o) = \left(\frac{\omega_o^4 \, n_i^4}{\pi \, c^4}\right)$, comes from the combined effect of frequency-dependent factors (namely areal density of states) and the energy (frequency) of the quanta being exchanged. The Bose factor, $F(B) = [n(\omega_o, T_1) - n(\omega_o, T_2)]$, provides the proportion of radiated quanta from each side.

We now define the other factors occurring in Eq. (9). The factor containing dipole terms is:

$$F(D) = \left(\frac{8\pi}{9}\right)|\mu_{mn}^{(1)}|^2 \, |\mu_{mn}^{(2)}|^2 \, N_1^a \, N_2^a \qquad (10)$$

The factor containing the material dielectrics is:

$$F(M) = \frac{\varepsilon_0^2}{|\varepsilon_0 + \varepsilon_1|^2 \, |\varepsilon_0 + \varepsilon_2|^2}, \quad \text{the material factor} \qquad (11)$$

The real part of the dielectric constant has only weak frequency dependence and is considered to be constant for our application.

The factor that gives the tunneling of evanescent waves across the gap (refractive index $n_g$) is:

$$T_{12}(\bar{\ell}) = \frac{1}{\left|1 + \frac{(|n_1|^2 + |n_g|^2)^2}{4|n_1|^2 \, |n_g|^2} \sinh^2(\bar{\ell})\right|} \qquad (12)$$

The geometric factor in Case 1 is:

$$G(\bar{\ell}) = 6\left(1/\bar{\ell}^{-4}\right)\left(\frac{W_{qw}}{\lambda_g}\right)^2 \text{ and } \bar{\ell} = \frac{\ell}{\lambda_g}, \quad W_{qw} \text{ width of quantum well} \qquad (13)$$

NUMERICAL ESTIMATE

For an approximate numerical estimate of the above equations, we choose the following values of the parameters involved:

$$E_{BG}(\text{band gap energy}) = 0.625 \, eV; \quad \omega_0 = 10^{15} \, \text{sec}^{-1}, \text{ and } n_1 = 3.5$$

$$F(\omega_o) = \left(\frac{\omega_o^4 \, n_i^4}{\pi \, c^4}\right) = \sim 6.4 \times 10^{19}/\text{cm}^4$$

$$\hbar = 10^{-27} \, ergs.\text{sec}, \quad \lambda_g = \lambda_o = 2 \times 10^{-4} \, cm = 2 \times 10^4 \, \text{Å}$$

$$|\mu_{mn}(1)|^2 = |\mu_{mn}(2)|^2 = \sim 2 \times 10^{-33} \, ergs \, cm^3$$

$$N_1^a = N_2^a = 2 \times 10^{16}/cm^2; \quad T_1 = 1000 \, K, \, T_2 = 300 \, K,$$

$$W_{qw} = 100 \, \text{Å}, \quad \ell = 1000 \, \text{Å}, \text{ and}$$

$$\bar{\ell} = (\ell/\lambda_g) = (1/20).$$

We have $\quad G(\bar{\ell}) = 24, \; T_{12}(\bar{\ell}) \approx 1, \; F(D) = 4.5 \times 10^{-33},$

and for $T_1 \gg T_2$, $\quad F(B) = e^{-\hbar\omega_0/K_B T_1} = 7 \times 10^{-4}$

as the factor $n(\omega_0, T_2)$ is too small to contribute.

Choosing $\varepsilon_0 = 1$, $\varepsilon_1' = \varepsilon_2' = 12$, $F(M) = 3.5 \times 10^{-5}$.

The net power density transferred between two adjacent 2-dimensional structures (from side 1 at $T_1$, to side 2 at $T_2$) is

$$P_{NF}(aa) = \sim 0.1 \, kW/cm^2$$

We have used plausible (and conservative) values for the various parameters involved. However, because of the high order in many cases (e.g., the fourth power), a change of some values by a factor of 2 can result in order-of-magnitude changes in the calculated power transfer.

Useful energy transfer has been shown to increase dramatically with decreasing gapwidth. As the absorber is removed to a distance from the source, the transverse waves can become the propagating photon mode. Figure 3 indicates the modeled gap dependence ($1/\ell^4$) of the transverse mode. Computer modeling[30] has confirmed these effects as well. The curvature at the bottom of the curve indicates the transition into the propagating mode. (The $n^2$ effect, with its $1/\ell^2$ dependence, may dominate in this transition region; or, it may be small compared to a greater enhancement resulting from the calculated proximity-resonance energy transfer.) Below the $10^{-7}$m gap region, the gap approaches the quantum well width selected and the $1/\ell^4$ dependence (from eq. 8) begins to roll over (not shown) and approaches the $1/\ell^2$ dependence of a thick well or a bulk semiconductor. Since the quantum-well result at high gap width is less than that of bulk materials, it will never cross over the bulk semiconductor curve.

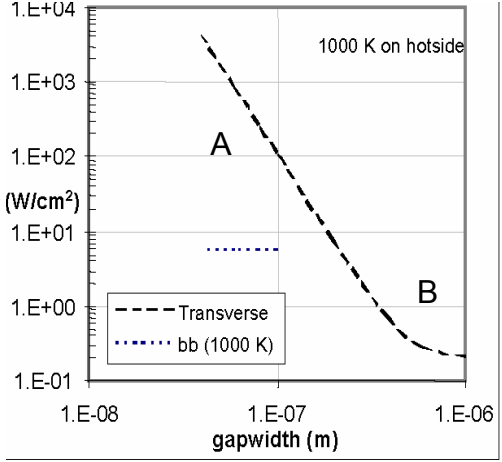

*Figure 3. The gap dependence of the power transferred. A) $1/\ell^4$ dependence region; B) transition region*

Included in the figure is the total radiating power from a 1000 K blackbody. At 0.1 μm, the sum of the proximity resonance energy transfer is >10 times that of the total blackbody radiation. Furthermore, this radiation may be nearly monochromatic rather than the broadband spectrum of the blackbody spectrum (most of which is not useful for photoconversion into electricity). Therefore, all the resonance-mode energy can be useful and significantly enhanced by the use of microgap structures. However, additional energy contributions from proximity coupling of free-carriers and absorption of propagating modes from outside of the quantum wells must be considered in the final accounting of the energy-transfer balance and utility.

COMPUTATIONAL ANALYSIS

A quantum mechanical model with great predictions[3] is necessary to interest some people (primarily physicists) in a new approach (in the present case, access to greatly beyond-black-body radiation levels). Other people (some engineers) are more pragmatic and have to be convinced by thermodynamic proofs[31] that conservation of energy is not being violated. One of the present authors was convinced early on of the $n^2$ effect by using, a simple geometrical argument, Snell's law, and total internal reflection. However, it was the computational analysis that convinced him that the "beyond $n^2$" effect was not wishful thinking. The fact that the QM predictions were supported by the computational analysis was comforting.

The computational analysis model developed at Draper Labs and MIT[32] is a particularly useful tool because it allows many different materials to be included in various layers. It is possible to model actual devices based on theoretical or, better yet, on experimentally-determined material parameters. The limitation of the model is that high-temperature optical parameters are not available for many materials. Thus, estimates had to be made for the emitter parameters. Nevertheless, some amazing confirmations and predictions can be made clear with this tool.

Figure 4 indicates the effect of the near-field coupling on power transfer between an emitter and a non-quantum-well TPV detector. The materials and structures represented here were optimized over a four-year development for both the spectrally-selective emitters[33,34] and the TPV devices.[35] Some aspects of this development are still proprietary. The figure shows the spectral power transferred from the emitter and absorbed in the detector. The peak power density absorbed from an emitter at a given temperature (e.g.

1000°C) and at two gap distances (0.1 and 100 μm) are seen to vary by more than a factor of 5. This calculated power enhancement is primarily a result of the $n^2$ effect; but, other factors are identified. The semiconductor emitter layer must be thin, so that free-carrier absorption at high temperatures does not introduce useless long-wave light.[36] Surface-plasmon-coupling spikes are identified. The selected antireflective-coating thickness is much more important for near-field coupling than for far-field emission. The high-refractive-index AR coatings, required for optimal near-field coupling, are not optimal for far-field emission.

With this type of emitter/TPV detector combination, the spectral efficiency (useable light energy divided by total light energy absorbed by the detector) can exceed 90%. With optimized TPV devices (based on actual devices, but modified for MTPV operation), thermal-to-electrical power conversion efficiencies in excess of 30% are predicted at emitter operating temperatures of ~900°C.

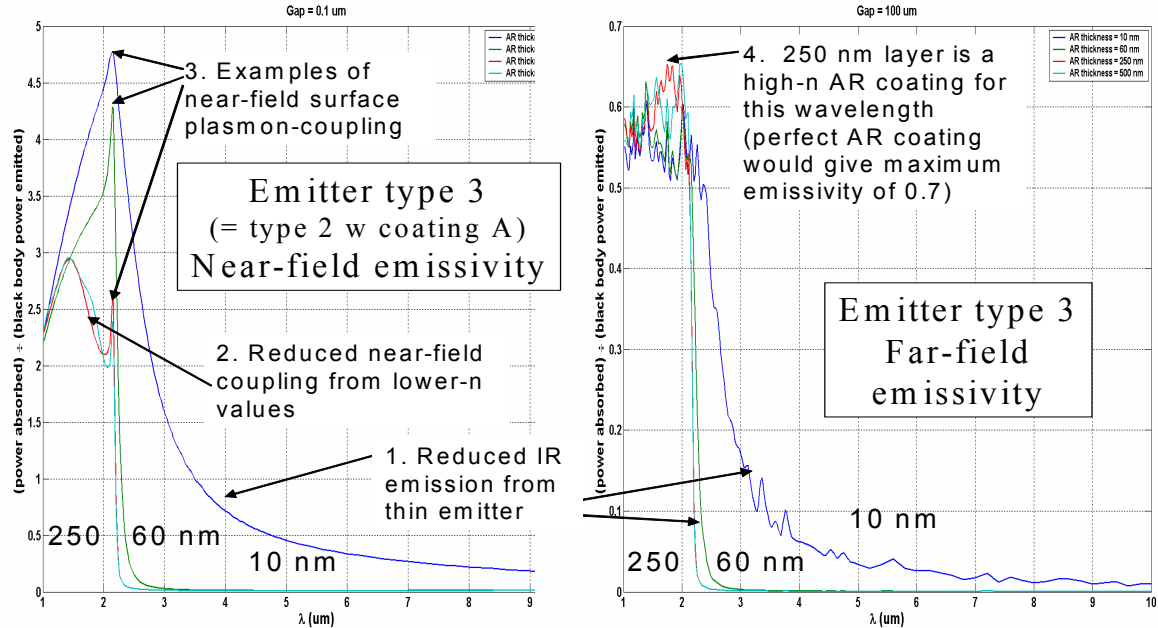

*Figure 4. Comparison of near-field (0.1 μm gap) and far-field (100 μm gap) emitter/detector spectral-power transfer. Four AR-coating thicknesses are compared (10, 40, 250, and 500 μm)*

Figure 5 demonstrates the effects of near-field resonant coupling (equation 8a) when the detector and emitter both have a common oscillator frequency. The emitter is the same in both systems; only the detector characteristic is changed. An order-of-magnitude enhancement of the resonant energy is seen with the addition of a resonant detector. Since conventional absorber materials were used in this model, rather than high-Q resonant quantum-well structures, the enhancement and total power transfer is only about 20% of that predicted by the numerical example in the prior section for a much-lower temperature. Nevertheless, predicted effects are seen qualitatively. For example, resonant AR-coating thickness (similar to the quantum well) is seen to be very important – up to a point (e.g., when AR-coating thickness (or QW width) approaches gap width, the enhancement is no longer quadratic as in Equ. (13)). Experimental work[22] to confirm the resonant coupling was never completed, in part because the theoretical and computational predictions were in sufficient agreement and the results were so compelling.

Quantum-well photodetectors, with a characteristic operating frequency, can be matched to the emitter resonance for such an efficiency-optimized system. The total power output might be less than for an un-optimized system; but, enhancement of light coupling at resonant, relative to non-resonant, frequencies would greatly reduce conversion losses. Therefore, the system efficiency would be very high. The trade-off between power output and system efficiency would determine the structure dimensions and materials chosen. It is clear that the "trade space" is quite large for such systems.

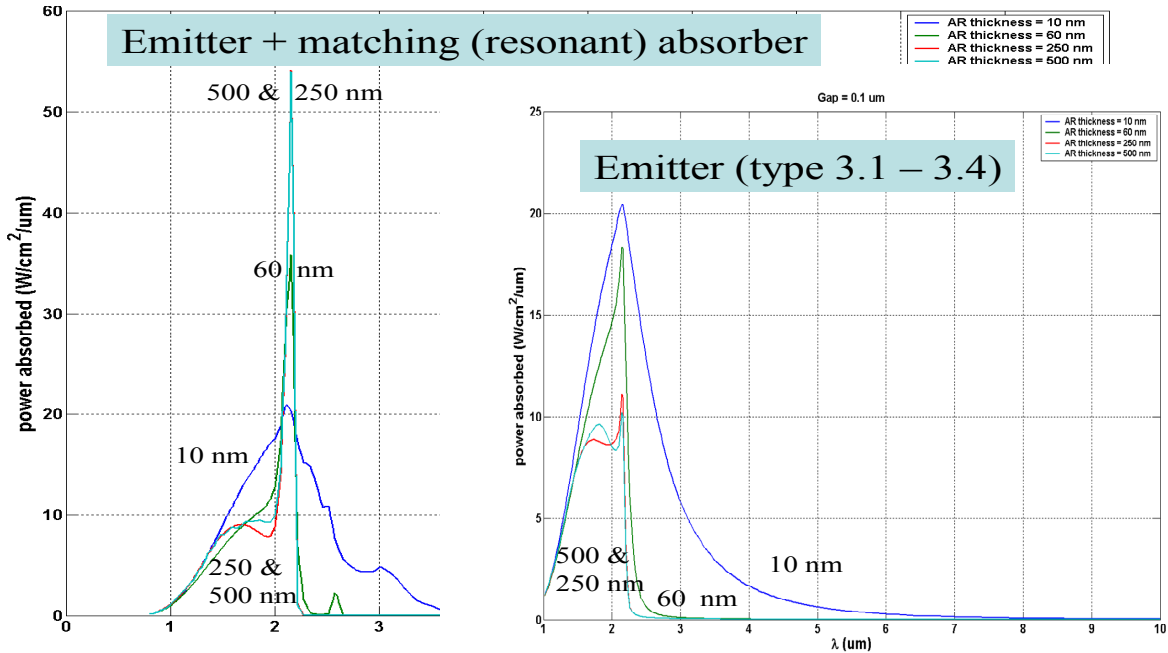

*Figure 5. Comparison of near-field (0.1 μm gap) emitter/detector spectral-power transfer when the detector is changed. Four AR-coating thicknesses are compared (10, 40, 250, and 500 μm)*

CONCLUDING REMARKS

A quantum-mechanical description of a problem that has been explored previously in a more classical setting has indicated that resonance effects can dominate energy transfer between a closely-spaced emitter and detector. Mulet et al.[37] indicated highly-enhanced radiative heat transfer between a small particle and a plane surface, which can support resonant-surface waves. Experimental confirmation of the resonant-proximity effects was a goal until the computational capabilities alone provided sufficient evidence.

The coupling between the atoms of the two slabs in this paper arises from the multipolar expansion of the Coulomb interaction between the charges (dipoles) of both sides mediated by the quanta of the electromagnetic field. In the near-field regime, dipole-dipole interaction is realized by the transfer of energy via transverse and longitudinal modes of the electromagnetic field.

The contribution of the longitudinal-photon mode (not included in the present calculation) turns out to be of the same order as that of the near-field transverse mode, which involves emission and absorption processes of real photons at resonance in the two-level situation envisioned. For the current case, the speed of light in the appropriate media is modified by the effective refractive index ($c \rightarrow c/n_e$). Thus, a factor $n_e^4$ occurs, which also enhances the power transfer. Further enhancement is possible if, instead of a vacuum gap, a dielectric gap is considered with $n_g > 1$. (Enhancement occurs for $n_g$ values up to that of the emitter and collector; however, we haven't figured out how to block the heat flow across a thin dielectric yet.)

The current quantum-mechanical resonant-coupling model confirms the predictions of both $n^2$ and "beyond $n^2$" enhancement, which removes a major limitation in thermophotovoltaics. Not only can the efficiency of TPV converters be increased, the emitter radiative efficiency can also be improved because energy that would otherwise be lost (in the form of long-wavelength light or heat) is recycled and selectively coupled into a resonant TPV converter. This selectivity further allows a reduction in the temperature of the emitter, while maintaining useful overall system efficiencies. A key feature to remember is that the new energy-transfer mechanism does not depend only on release of the blackbody radiation trapped within the emitter (as does the classical $n^2$ effect). The additional energy source is the non-propagating photon modes that are normally dissipated in self-excitation of the emitter atoms and in resonance-coupling effects. This means that the blackbody law of power emission (which pertains only to the propagating modes) is not violated. We can <u>not</u> get more power out than we put in. However, we <u>can</u> extract energy more rapidly and more selectively at any emitter temperature. Therefore, with microgap coupling and a given thermal-energy input, the emitter can be kept at a lower temperature and still operate at a higher efficiency than previously possible. This is important in that it opens the possibility of selecting emitter materials and structures that would not survive higher temperatures.

ACKNOWLEDGEMENTS

Part of this paper is based on work at MIT and Draper Labs, supported by The Charles Stark Draper Laboratory, Inc., Cambridge, MA  02139 and is funded in part by the Science for Humanity Trust, Bangalore, 560012, India, the Science for Humanity Trust, Inc, Tucker, GA, USA, and the Indian National Science Academy (for KPS). Special thanks are due Mark Weinberg and Eric Brown of Draper Labs for setting up the computational model used here and for extending its capabilities and making it so user friendly respectively.

quantum-well model, we are assuming a dielectric "window" between the two slabs, rather than a vacuum gap (which would generate evanescent modes at the interface where the refractive index changes abruptly). Tunneling across the two interfaces of a vacuum gap leads to the classical $n^2$ effect. It is included later (in Equ. (9)) and described in Equ. (12) as a factor $T_{12}(\bar{\ell})$ in the proximity function.